# Optimally Setting RMS Energy Bands for Detecting Oscillations

Chetan Mishra, *Member, IEEE*, Luigi Vanfretti, *Senior Member, IEEE,* Bikal Pudasaini, *Member, IEEE*, Jaime Delaree Jr., *Member, IEEE* and Kevin D. Jones, *Member, IEEE*

***Abstract*—Oscillation monitors have root-mean-square (RMS) energy detectors at their core. These detect oscillation in a predefined frequency band by comparing the signal power in that band with a threshold (usually set empirically). A key design decision is the choice of frequency bands. Traditionally, for monitoring the stability of wide-area electromechanical modes, finer bands were set below 1 Hz, while higher frequencies were covered by wider bands. However, this approach can yield poor detection in modern power systems with fast and largely unknown local dynamics. This letter proposes an algorithm to set frequency bands from the ambient noise spectrum to maximize the average detection performance.**

## I. Introduction

Synchrophasors are critical for tracking power grid dynamics, particularly in the modern grid with a growing number of high-frequency oscillations due to power electronics and a lack of accurate models to analyze them. The real-time detection of electromechanical and forced oscillations is a key function of wide-area monitoring systems. Among automated monitoring tools, the root-mean-square (RMS) energy detector has become an industry standard due to its model-free architecture and low computational cost. The work in [1] championed the use of signal processing techniques on continuous grid data to characterize system dynamics, laying the foundation for modern RMS energy filters. The work in [2] proposed the standard framework for RMS energy detectors, which isolates the band of interest by filtering, followed by squaring and smoothing, yielding a computationally lightweight index reflecting the localized energy of an oscillation. However, early field deployments required intensive tuning, particularly of the alarm threshold. Recent work [3], [4] developed multiple statistical thresholding approaches relying on characterization of ambient noise to minimize false alarms, the more recent of which also adds amplitude-based thresholds to filter out low-impact events.

Traditionally, RMS energy detectors were mainly used to track the stability of known dynamics, with an emphasis on slow inter-area modes. This is reflected in the bands typically set for frequency signals: a narrow 0.15–1 Hz band followed by wide 1–5 Hz (for local modes) and 5–30 Hz for inverter-based resource (IBR) and other fast dynamics [2]. With the growth of IBRs [5], high-frequency dynamics have become common, altering the ambient noise characteristics. Therefore, the choice of RMS energy bands must be adapted to capture these oscillations effectively. To address this gap, this letter proposes a method to optimally set RMS energy detector bands from the ambient noise spectrum to improve oscillation detection.

## II. Main Result

### A. Characterization of RMS Energy

Given a segment of length $N$ of ambient data $x_n$, let $I^+$ be the set of indices on the discrete Fourier transform (DFT) grid corresponding to sorted positive frequencies. Denote the periodogram [6] at $i^{th}$ frequency by $\hat{P}_i^N$. A well-known statistical characterization of Gaussian processes (as assumed for ambient data) with power spectral density (PSD) $S_i$ is $\lim_{N\to\infty} \hat{P}_i^N \sim \frac{S_i}{2} z_i$ where $z_i \sim \chi_2^2$ are i.i.d. The RMS energy $E_{RMS}(B)$ in a frequency band defined by a set of DFT frequency indices $B \subseteq I^+$ can be expressed as [3],

$$\frac{N}{2} \times E_{RMS}^2(B)|_{ambient} = \sum_{i\in B} \hat{P}_i^N = \sum_{i\in B} \frac{S_i}{2} z_i \quad (1)$$

During an oscillation of amplitude $A_o$ at the frequency with index $i_o$, it can be shown that

$$\frac{N}{2} \times E_{RMS}^2(B)|_{A_o>0,i_o} = \sum_{i\in B} \frac{S_i}{2} z_i + \epsilon_{A_o,i_o} \times 1(i_o \in B) \quad (2)$$

where $O(N)$ denotes a term of the order of the window length $N$, with the stochastic term $\epsilon_{A_o,i_o}$ being Gaussian distributed, with mean and variance both of the order $A_o^2 O(N) > 0$, combining the deterministic oscillation power and the cross term between the ambient noise and the oscillation component. Moreover, $\epsilon_{A_o,i_o}$ is independent of $z_i \forall i \neq i_o$. Hereafter, we also write $E_{RMS}^2(B)|_{ambient}$ as $E_{RMS}^2(B)|_{A_o=0}$.

RMS energy detectors rely on the RMS energy in a band increasing during an oscillation relative to its ambient value. Comparing (1) and (2), the only additional term during an oscillation is $\epsilon_{A_o,i_o}$ which must therefore be positive for the RMS energy to increase, a condition that does not always hold.

**Theorem 1:** For a large window length $N$, $\epsilon_{A_o,i_o} > 0$ with high probability.

**Proof:** Given $\epsilon_{A_o,i_o} \sim \mathcal{N}\left(\mu = A_o^2 O(N),\ \sigma^2 = A_o^2 O(N)\right)$. By Cantelli's inequality applied to the left tail, for every negative constant $\beta^-$,

$$\Pr\left(\epsilon_{A_o,i_o} - \mu < \beta^- - \mu\right) \leq \frac{1}{1 + \frac{(\beta^- - \mu)^2}{\sigma^2}} = \frac{1}{1 + \frac{\left(\beta^- - A_o^2 O(N)\right)^2}{A_o^2 O(N)}}$$

Therefore $\lim_{N\to\infty} \Pr\left(\epsilon_{A_o,i_o} < \beta^-\right) \to 0 \forall \beta^-$.

Chetan Mishra, Bikal Pudasaini, Jaime Delaree Jr. and Kevin D. Jones are with Dominion Energy, Richmond, VA 23060 USA (e-mail: chetan.mishra@dominionenergy.com). Luigi Vanfretti is with Rensselaer Polytechnic Institute, Troy, NY, USA (e-mail:vanfrl@rpi.edu )

This condition, i.e., $\epsilon_{A_o,i_o} > 0$ with high probability, is what allows standard detectors of the form $E^2_{RMS}(B) > threshold$ to be valid.

We now evaluate the impact of the band choice $B$ on detecting an oscillation at $i_o$ irrespective of the specific thresholding strategy [3], [4]. Statistically, a detector of the form $E^2_{RMS}(B) > threshold$ is expected to perform well if the distributions of $E^2_{RMS}(B)|_{A_o=0}$ and $E^2_{RMS}(B)|_{A_o>0,i_o}$ are well separated, with the latter to the right of the former. A popular metric to quantify this is the area under the receiver operating characteristic (ROC) curve (AUC), defined as

$$AUC(B, A_o, i_o) = \Pr\left(E^2_{RMS}(B)|_{A_o>0,i_o} > E^2_{RMS}(B)|_{A_o=0}\right) \quad (3)$$

Substituting (1) and (2) into (3),

$$\Rightarrow \Pr\left(\underbrace{\sum_{i\in B} \frac{S_i}{2}\left(z_{i,1} - z_{i,0}\right)}_{Z_B} > -\epsilon_{A_o,i_o} \times 1(i_o \in B)\right) \quad (4)$$

where $z_{i,0}, z_{i,1} \sim \chi^2_2$ are i.i.d. An optimal $B$ is one that maximizes the AUC. If $i_o \notin B$, the right-hand side of (4) is 0 and the AUC becomes $\Pr(Z_B > 0) = 0.5$ since $Z_B = \sum_{i\in B} \frac{S_i}{2}\left(z_{i,1} - z_{i,0}\right)$ is symmetrically distributed about zero. Hence, the optimal $B$ must contain $i_o$.

### *B. Multi-Band Setting Formulation and Proposed Solution*

RMS energy detectors aim to detect oscillations at all frequencies; however, a detector for each frequency is computationally prohibitive in practice. Instead, typically 4–5 contiguous, non-overlapping bands are used to cover the whole range $I^+$. Let $p_{i_o}$ be the probability that an oscillation occurs at the $i_o^{th}$ frequency. The problem of optimally setting $M$ bands can then be formulated as

$$\max_{b_0 \dots b_M} \sum_m \sum_{i_o \in B_m} p_{i_o} \times AUC(B_m, A_o, i_o) \quad (5)$$
$$s.t.\ B_m = \{i \in I^+ | b_{m-1} < i \le b_m\}\ and \bigcup_{m=1:M} B_m = I^+\ and\ b_0 < b_1 < \dots < b_M$$

Here, $b_0 < b_1 < \dots < b_M$ are boundary indices of the contiguous, non-overlapping bands. However, $AUC$ is a function of the oscillation amplitude $A_o$, which is unknown. Moreover, the aim is to isolate the effect of the band choice $B$. For a large window length $N$, Theorem 1 guarantees that for any small positive constant $t > 0$, the random variable $\epsilon_{A_o,i_o}$ exceeds $t$ with high probability. Consequently, $\Pr(Z_B > -t)$ serves as an asymptotic lower bound for the AUC in (4). Furthermore, $\Pr(Z_B > -t)$ can be simplified when the band $B$ contains a sufficiently large number of frequencies, as is typically the case when a few bands cover the whole range, using the central limit theorem [7] and a first-order approximation (small t)

$$\Pr(Z_B > -t) = \Phi\left(\frac{t}{\sqrt{VAR(Z_B)}}\right) \approx 0.5 + \frac{1}{\sqrt{2\pi}} \frac{t}{\sqrt{VAR(Z_B)}} \quad (6)$$

where $VAR(Z_B) = 2\sum_{i\in B} S_i^2$ and $\Phi$ is the cumulative distribution function of the standard normal distribution $\mathcal{N}(0,1)$. Finally, replacing $AUC(B_m, A_o, i_o)$ in (5) by $\Pr(Z_B > -t)$ simplifies the optimization problem to

$$\max_{b_0 \dots b_M} \sum_m \frac{\sum_{i_o \in B_m} p_{i_o}}{\sqrt{\sum_{k\in B_m} S_k^2}} \quad (7)$$
$$s.t. constraints\ from\ (5)$$

In (7), the numerator is the probability of an oscillation occurring in a band, while the denominator quantifies the difficulty of detection in that band, as it is proportional to the standard deviation of the ambient RMS energy. This partitioning problem can be solved using dynamic programming [8]. In summary, the proposed method sets the RMS energy bands as a function of the PSD $S$ to maximize the separation between the resulting RMS energy distributions with and without an oscillation, leading to more effective detectors.

## III. Results

Consider first a simple case with a single well-damped mode (10% damping) at 0.5 Hz, whose true PSD is shown in Fig. 1, mimicking a system with only conventional slow electromechanical dynamics. Phasor measurement unit (PMU) data are reported at $f_s = 30$ Hz with a window length of 60 s ($N = 30 \times 60 = 1800 \Rightarrow |I^+| \approx 900$) used in the RMS energy detector. The goal is to set $M = 2$ detection bands characterized by a single breakpoint $b_1$, $b_0 = 0$ and $b_2 = 899$, to maximize the probability of detecting oscillations.

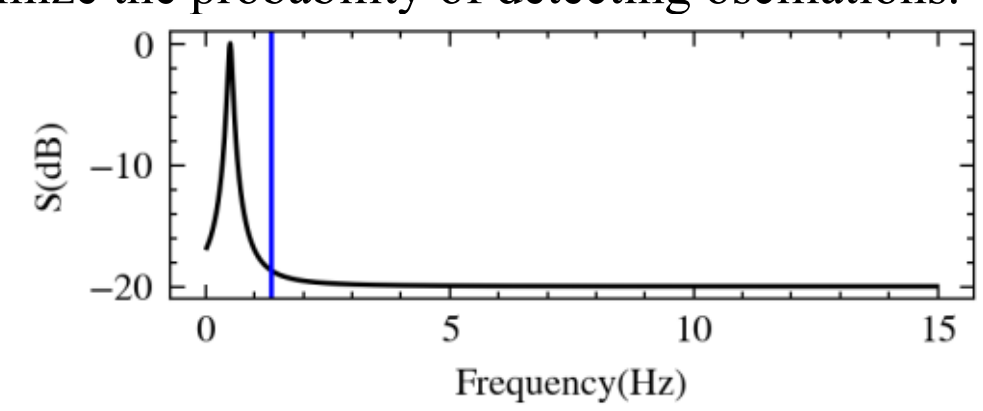

**Fig. 1. Single-mode case, $M = 2$.**

First, oscillations are assumed equally likely at all frequencies, i.e., $p_{i_o} = \frac{1}{|I^+|}$. The optimal value of $b_1$ is shown in blue in Fig. 1 at 1.35 Hz, obtained by solving (7). To empirically evaluate the optimality of this solution, we generate two sets of 1000 independent 60-s ambient data windows with PSD $S$ as plotted in black in Fig. 1. We create $|I^+|$ copies of the second set, where the $i_o^{th}$ copy has a fixed $A_o$-amplitude oscillation added at the $i_o$ frequency. For each candidate solution ($b_1$), we calculate the RMS energies of the bands $B_1 = (0, b_1]$ and $B_2 = (b_1, |I|^+]$ to empirically estimate $AUC(B_m, A_o, i_o) \forall m, i_o$ and hence the objective function in (5). In Fig. 2, this actual objective is compared with the approximate one in (7) for different values of $A_o$.

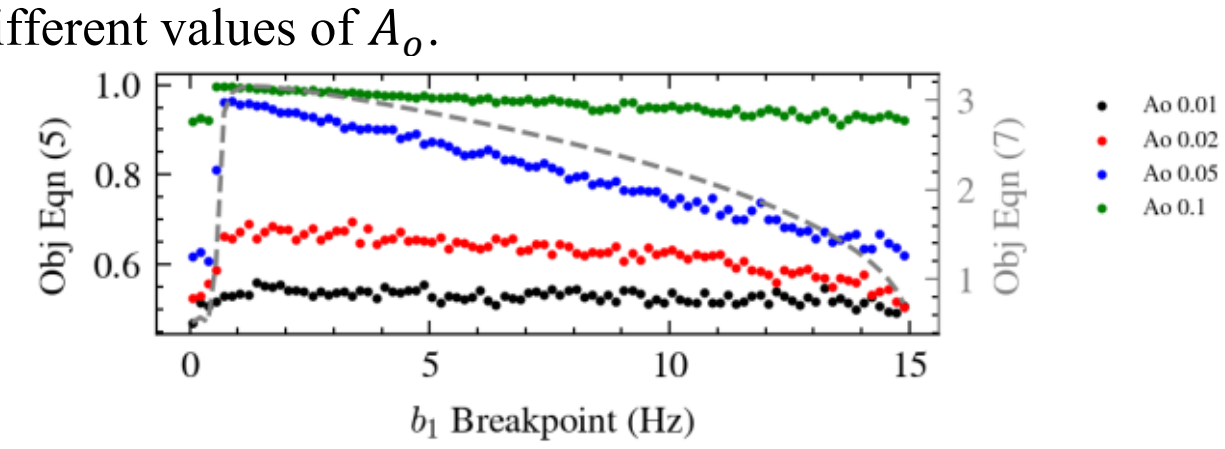

**Fig. 2. Actual objective (5) vs. approximate objective (7).**

The simplified expression in (7) captures the shape of the actual objective well, while being much simpler to evaluate. The downward trend is explained as follows. The range below 1 Hz has high PSD values; therefore, the RMS energy of the left band has a much higher variance $VAR(Z_B)$ (see (6)) and, consequently, poorer detection performance (for the same $A_o$) than the right band. Widening the left band (increasing $b_1$) places more frequencies in this poorly performing band, which explains the decline of the objective function.

To demonstrate how an optimal band setting (high average AUC across all frequencies) translates into improved detection accuracy, we benchmark against the statistical thresholding approach in [3] for detecting an oscillation at an arbitrarily chosen frequency of 1.6 Hz. Under ambient conditions, the empirical distribution of the RMS energy of the band containing the oscillation (from the previously generated data) is plotted in black for two cases, the optimal partition at 1.35 Hz and a suboptimal partition at 1.85 Hz, in Fig. 3. From these distributions, a threshold corresponding to a significance level $\alpha = 0.05$ is computed as in [3] to limit the Type I error (nuisance alarms) in each case. Next, oscillations of varying amplitudes $A_o$ are added at 1.6 Hz, and the resulting RMS energy distributions are also shown in Fig. 3. The detection accuracy is the probability that the RMS energy exceeds the threshold. For a small amplitude of 0.01, detection is inherently limited because the ambient and oscillation distributions largely overlap; band optimization therefore offers little benefit, and the detection probability remains below 0.1. For larger amplitudes, however, the optimal partition substantially improves detection accuracy, in some cases more than doubling it. This also explains the trend of the objective functions with respect to $A_o$ seen in Fig. 2: the benefit of optimization can be negligible for very small amplitudes because of the poor signal-to-noise ratio.

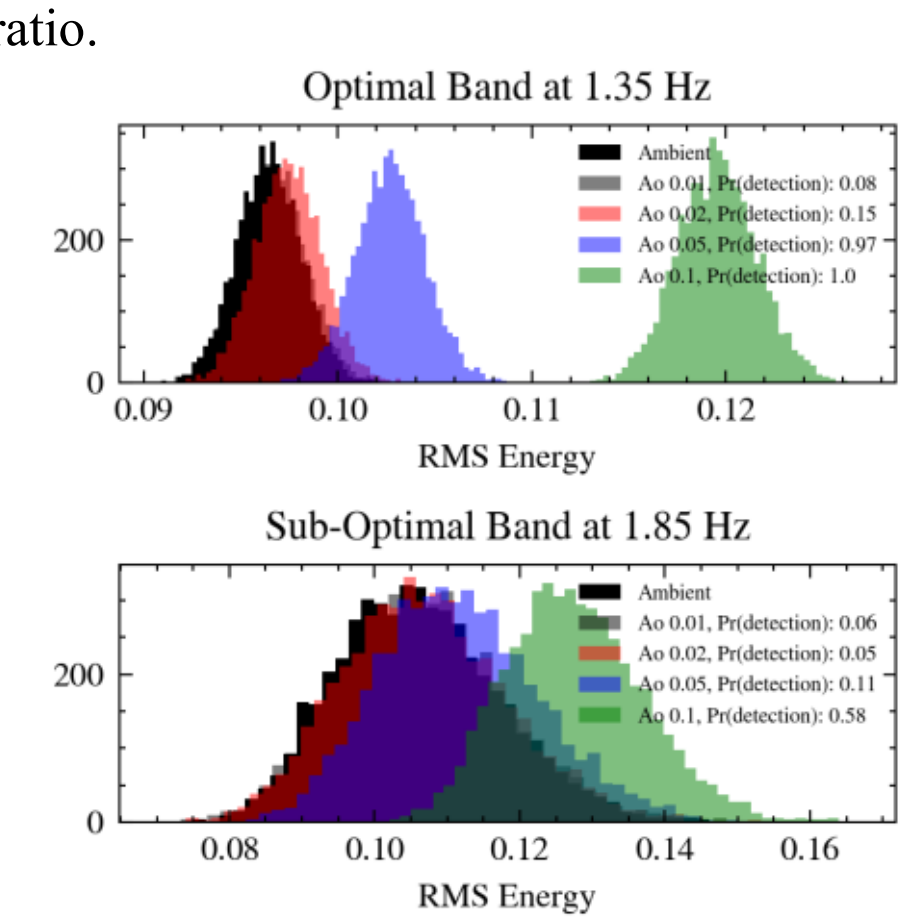

**Fig. 3. Oscillation at 1.6 Hz, detection at $\alpha = 0.05$.**

When all frequencies are treated as equally likely to oscillate ($p_{i_o} = constant$), the optimization keeps the left band as narrow as possible to confine the problematic high-$S$ region. In practice, however, physical systems produce dynamics only on limited time scales, regardless of the data sampling rate. For the previous example, assume that the probability of high-frequency oscillations decays exponentially with frequency, i.e., $p_{i_o} \propto \exp(-\lambda i_o)$ as shown in red in the left plot of Fig. 4. The resulting optimal bands for $M = 4$ are shown for the original uniform $p_{i_o}$ (blue) and the exponential one (red). For the former, as expected, the problematic leftmost band is unchanged (as in the $M = 2$ case in Fig. 1) while the right band is split further to improve detection in the high-frequency range, as shown by the blue partitions in Fig. 4. In contrast, since the exponential distribution down-weights that frequency range (left subplot of Fig. 4), the algorithm focuses on improving performance in the 0–2 Hz range, while a single wide band covers the range above 2 Hz.

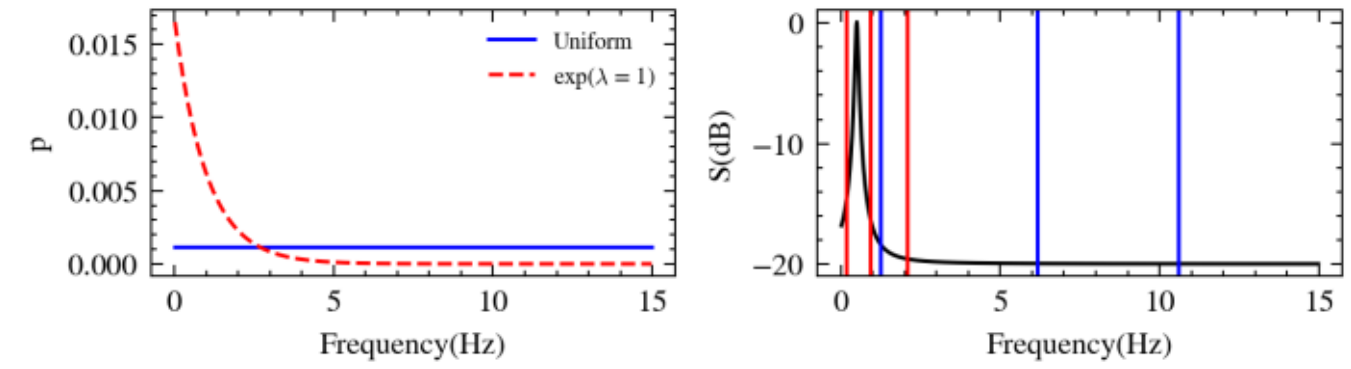

**Fig. 4. Effect of p, M = 4.**

Finally, a good rule of thumb for the likely oscillation frequency range, as used above, may not always be available. In that case, the PSD $S$ itself can serve as a guide by choosing $p_{i_o} \propto S_{i_o}$. This places more weight on detection performance in ranges with high PSD, i.e., with more dynamics and a higher propensity to oscillate. Insight into the resulting band-level detection performance is gained from the AUC surrogate given by the inverse of the standard deviation of the band RMS energy $\sigma(Z_{B_m}) \propto \sqrt{\sum_{k \in B_m} S_k^2}$ in (6). Fig. 5 shows this surrogate in red for each band for a complex-shaped $S$, where using $p \propto S$ improves detection performance (lower $\sigma(Z_{B_m})$) near two prominent dynamics below 10 Hz, which are largely ignored under a uniform $p$. For context, reducing $\sigma(Z_{B_m})$ to one-fourth in a frequency range roughly allows oscillations of half the original amplitude to be detected with the same accuracy.

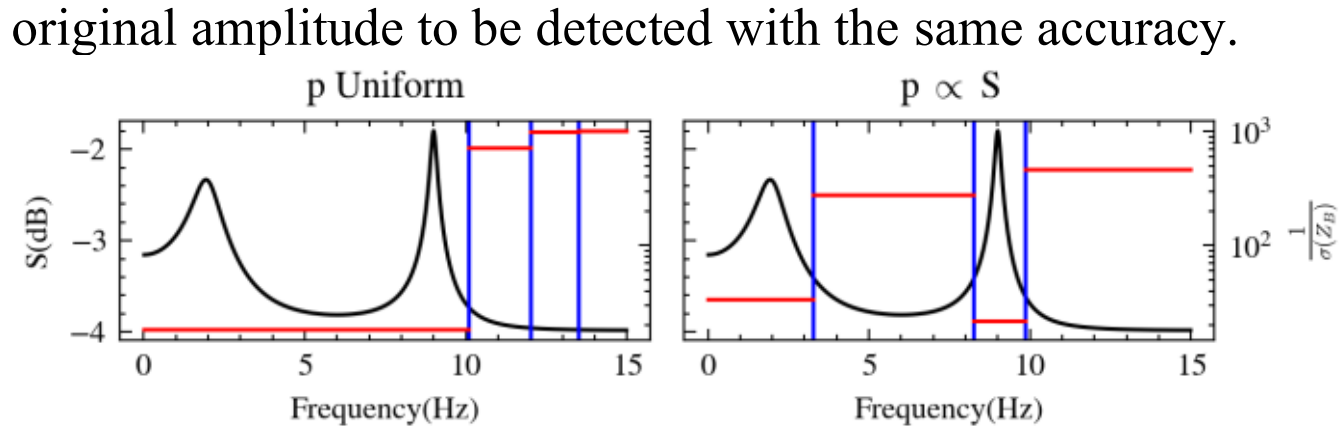

**Fig. 5. Complex PSD and insight from the approximate AUC.**

This demonstrates that the framework can easily incorporate real-world constraints and preferences.